\documentclass[fleqn,twoside]{article}
\usepackage{espcrc2}

\usepackage{graphicx}

\usepackage[figuresright]{rotating}

\title{The Timing Counter of the MEG experiment: calibration and performance}

\author{
       P.~W.~Cattaneo~ 
        \address[pvinfn]{INFN Pavia, Via Bassi 6, I-27100, Pavia, Italy.}~
        \thanks{Corresponding author: Paolo.Cattaneo@pv.infn.it
        }, %
       M.~De~Gerone 
        \address[geuni]{INFN and Universit\'a di Genova, Dipartimento di Fisica, 
          Via Dodecaneso 33, I-16146, Genova, Italy
        }, %
       S.~Dussoni 
        \addressmark[geuni],
       F.~Gatti 
        \addressmark[geuni],
       M.~Rossella 
        \addressmark[pvinfn],
       Y.~Uchiyama~ 
        \address[tokyo]{ICEPP, The University of Tokyo, 7-3-1 Hongo, Bunkyo-ku, Tokyo 113-0033, Japan}, %
       R.~Valle 
        \addressmark[geuni]
       }
       
\begin{document}

\begin{abstract}
The MEG detector is designed to test Lepton Flavor Violation in the $\mu^+\rightarrow e^+\gamma$ 
decay down to a Branching Ratio of a few $10^{-13}$.
The decay topology consists in the coincident emission of a monochromatic photon 
in direction opposite to a monochromatic positron. A precise measurement of the relative
time $t_{e^+\gamma}$ is crucial to suppress the background.
The Timing Counter (TC) is designed to precisely measure the time of arrival of the $e^+$
and to provide information to the trigger system.
It consists of two sectors up and down stream the decay target, each consisting of two layers.
The outer one made of scintillating bars and the inner one of scintillating fibers. Their design 
criteria and performances are described.

\end{abstract}

\maketitle

\section{The MEG experiment }

MEG is an experiment designed to improve significantly the limits on the Branching Ratio 
(BR) of the $\mu^+ \rightarrow e^+ \gamma$ decay \cite{Meg99}-\cite{Meg02}. This decay channel 
is strongly suppressed in the Standard Model but is allowed to a measurable level in 
many alternative theories \cite{Cattaneolfv}.\\
The topology of this decay is very simple, consisting of monochromatic $e^+$ 
and $\gamma$ with a common space time origin, 
with energy close to half the $\mu^+$ mass, moving in opposite directions.
These features can be summarized as
\begin{itemize}
\item $E(e^+) = E_{e^+}^{sig} \approx m_{\mu^+}/2$
\item $E(\gamma) = E_{\gamma}^{sig} \approx m_{\mu^+}/2$
\item $T_{e^+\gamma} = T_{e^+} - T_{\gamma} = 0$
\item $\overline v_{e^+} = \overline v_{\gamma} $
\item $\Theta_{e^+\gamma} = \Theta_{e^+} - \Theta_{\gamma} = \pi $
\end{itemize}
where $\overline v$ is the origin space vertex, $T$ the origin time, $\Theta$ the 
relative polar angle and $E$ the energy.\\
The main difficulty in this measurement is that, in order to collect the high 
statistic required to improve the existing BR in a few years, a very high rate 
of $\mu^+$ must decay on the target: $\approx 3\,10^7 Hz$ is the design value.
The Michel decay $\mu^+ \rightarrow e^+ \nu \overline \nu$ 
produces $e^+$ with an energy up to $E_{e^+}^{sig}$,
therefore high resolution energy measurement is required to reduce the Michel $e^+$
contamination.\\
The $\gamma$ background originates from radiative decay (RD) $\mu^+ \rightarrow e^+ \gamma 
\nu \overline \nu$, $e^+e^-$ annihilation in flight and $e^+$ bremsstrahlung. The 
kinematic limit is $ E_{\gamma^+}^{sig}$ and therefore background suppression
needs high resolution measurement of $E(\gamma)$.\\
The $\gamma$ and $e^+$ directions and the $e^+$ vertex need to be measured 
with high precision to discriminate combinatorial background from random 
association of $e^+$ and $\gamma$ from different $\mu^+$ decays.
Precise measurements of $T_{e^+}$ and $ T_{\gamma}$ 
reduce the combinatorial background from different $\mu^+$ decays.\\
The photon detector is a Liquid Xenon Calorimeter (LXe) exploiting scintillation 
light \cite{Sawada09} that provides a measurement of the photon energy, position 
and timing.\\
The positron is observed by a spectrometer having at the core a set of 16 low mass 
Drift Chambers (DCH) embedded in a high magnetic field (${\cal O}(1 T)$) with a gradient
along the z axis to bend the positron with a radius weakly dependent on the emission
polar angle. The DCH measures the $e^+$ momentum and position 
but is unable to deliver trigger information and to provide precise 
timing \cite{Cattaneo09}.
The trigger and timing information on the $e^+$ are delivered by
the Timing Counter, that is described in detail in the following.

\section{The Timing Counter: the design}

The TC is required to cover the solid angle opposite to the LXe 
providing high efficiency for $e^+$ detection. This requirement is 
satisfied dividing
the TC in two modules, called sectors, placed symmetrically with respect 
to the decay target. Each sector must provide a precise measurement of the 
$e^+$ timing both at the trigger and analysis levels (${\cal O}(100ps)$). 
Furthermore it provides a measurement of the crossing 
point both for the trigger, requiring the $e^+$ and $\gamma$ moving in
opposite direction, and for the analysis, DCH-TC match and determination of the $e^+$
track path length.\\
These requirements are met with a two layer detector: the outer one 
measures the transverse coordinate ($\phi$ with respect to the beam direction) 
and provides in addition time information for trigger and analysis.
The inner one measures the longitudinal coordinate ($z$ along the beam direction)
providing trigger and analysis information. 

\subsection{The Longitudinal Timing Counter}

\begin{figure}[htb]
\begin{center}
\includegraphics[width=6cm,angle=0]{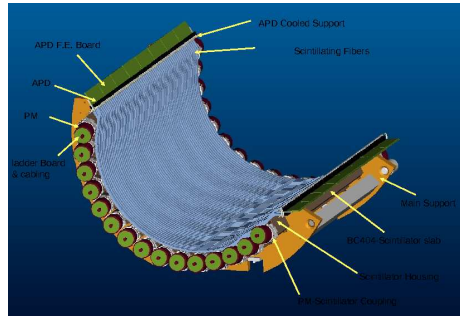}
\end{center}
\caption{The MEG timing counter }
\label{MEGTC}
\end{figure}

The outer layer consists of 15 scintillator bars located along the z-axis at fixed radius 
($\approx 30\,cm$) in a barrel-like array with 10.5$^\circ$ gap (see Fig.\ref{MEGTC}).
This configuration has high acceptance for $\mu^+\rightarrow e^+\gamma$ decay
with momentum $p_{e^+} = 52.83\,MeV/c$ and reduced acceptance for Michel events.
The number of bars is matched to the number of DCH and to the trigger requirement for 
selecting collinear $e^+-\gamma$.
The bars have a square section with edge $4.0\,cm$ and length $80\,cm$ and are read by 
fine-mesh PMTs adequate for use in high intensity magnetic field.
The criteria leading to the choice of these parameters are explained in \cite{Dussoni10}.\\
The signals from the PMTs are processed by a Double Threshold Discriminator (DTD) with 
a low threshold to reduce Time Walk effect and a high threshold to remove background events.
When the DTD is fired, it delivers a NIM signal.\\
The PMT and the NIM signals are read by the Domino Ring Sampler a custom designed 
digitizer operating in MEG at frequency up to 2GHz \cite{Ritt2007}. The full waveforms are 
stored so that the time can be extracted offline with optimized algorithms.

\subsection{The Transverse Timing Counter}

The inner layer is made of 256 $5 \times 5 mm^2$ scintillating fibers providing trigger 
and analysis information on the $z$ coordinate \cite{DeGerone09} 
(see Fig.\ref{MEGTC}). Each fiber is read out by APDs (Avalanche Photo Diode) that allow 
operations in high intensity transverse magnetic field with a gain $\approx 10^3$.\\
The analog signal from the APD are filtered, 16 of them are summed and made available to 
the trigger. In parallel, the signals are discriminated and the digital hitmap is made 
available for the analysis.

\section{The Timing Counter: calibration and performances}

\begin{figure}[htb]
\begin{center}
\includegraphics[width=8cm]{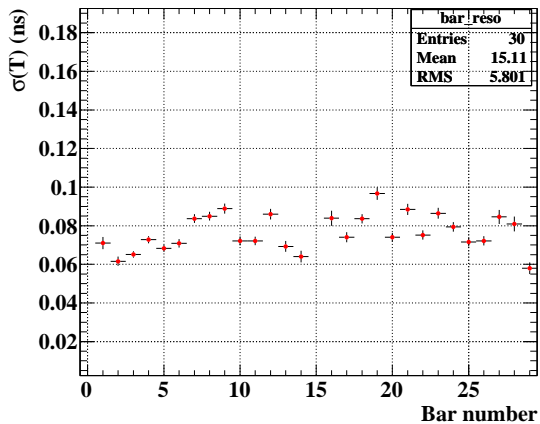}
\end{center}
\caption{The single bar time resolution using two-bar coincidence.}
\label{barreso}
\end{figure}

The longitudinal layers started taking data in 2006 and have been active during the 
following years. To achieve the design resolution, it turned 
out necessary to calibrate carefully the detector.\\
The calibration tools naturally available are the large flux of Michel $e^+$
from $\mu^+$ decays and the cosmic rays. There are two other calibration sources
requiring a dedicated set up. A $\pi^-$ beam impinging on a liquid $H$ target 
produces through charge exchange reaction Dalitz decays $\pi^0\rightarrow \gamma e^+e^-$. 
Protons of kinetic energy $T\approx 1 MeV$ from a custom Cockroft-Walton accelerator 
produces on a Boron target the nuclear reaction $^{11}_5B(p,2\gamma)^{12}_6C$.\\
Important calibration items are the Time Walk (TW) terms that account for the 
amplitude dependence of the timing measurement. They are measured using $e^+$ hitting two or three
bars. The same events are used to measure the TC time resolution.\\
Other crucial calibration is the measurement of timing offset of the different PMTs 
to allow combining results from different part of the TC. This calibration is obtained
with the Dalitz or Boron events, that have two particles emitted at the same time, 
using the LXe as reference.\\
The single bar time resolutions with TW correction and offset subtracted
are shown in Fig.\ref{barreso}.\\
The transverse layer worked only partially due to problems with the digital readout and 
excess noise, that prevented an efficient use of the detector. With the data available, it
was possible to verify that the tracks reconstructed from DCH match the hits on both
transverse and longitudinal layers reducing the combinatorial background and delivering 
improved information on the $e^+$ track length.

\section{Conclusions}

The TC of MEG has performed as high resolution timing detector according to the expectations playing
a crucial role in the MEG trigger system. The transverse layer 
met unexpected problems with part of the digital readout, but the working part delivered satisfactory
results. For the 2010 run, both layers are expected to be fully operational.

\end{document}